\documentclass[pra,twocolumn,superscriptaddress,showpacs,preprintnumbers,amsmath,amssymb]{revtex4-1}
\usepackage{graphicx}% Include figure files
\usepackage{dcolumn}% Alig table columns on decimal point
\usepackage{bm}% bold math
\usepackage{enumerate}
\usepackage{mathdots}
\usepackage{epstopdf}
\usepackage{color}
\usepackage[table]{xcolor}
\begin{document}
\title{An analytical double-unitary-transformation approach for strongly and periodically driven three-level systems}

\author{Yingying Han}
\affiliation{School of Physics and Technology, Wuhan University, Wuhan, Hubei 430072, China}
\affiliation{Shenzhen Institute for Quantum Science and Engineering,
Southern University of Science and Technology, Shenzhen 518055, China}

\author{Xiao-Qing Luo}
\affiliation{Hunan Province Key Laboratory for Ultra-Fast Micro/Nano Technology and Advanced Laser Manufacture, School of Electrical Engineering, University of South China, Hengyang 421001, China}

\author{Tie-Fu Li}
\affiliation{Institute of Microelectronics, Tsinghua University, Beijing 100084, China}
\affiliation{Beijing Academy of Quantum Information Sciences, Beijing 100193, China}

\author{Wenxian Zhang}
\email[Corresponding email: ]{wxzhang@whu.edu.cn}
\affiliation{School of Physics and Technology, Wuhan University, Wuhan, Hubei 430072, China}

\date{\today}

\begin{abstract}
Floquet theory combined with the generalized Van Vleck nearly degenerate perturbation theory, has been widely employed for studying various two-level systems that are driven by external fields via the time-dependent longitudinal (i.e., diagonal) couplings. However, three-level systems strongly driven by the time-dependent transverse (i.e., off-diagonal) couplings have rarely been investigated, due to the breakdown of the traditional rotating wave approximation. Meanwhile, the conventional perturbation theory is not directly applicable, since a small parameter for the perturbed part is no longer apparent. Here we develop a double-unitary-transformation approach to deal with the periodically modulated and strongly driven systems, where the time-dependent Hamiltonian has large off-diagonal elements. The first unitary transformation converts the strong off-diagonal elements to the diagonal ones, and the second enables us to harness the generalized Van Vleck perturbation theory to deal with the transformed Floquet matrix and also allows us to reduce the infinite-dimensional Floquet Hamiltonian to a finite effective one. For a strongly modulated three-level system, with the combination of the Floquet theory and the transformed generalized Van Vleck perturbation theory, we obtain analytical results of the system, which agree well with exact numerical solutions. This method offers a useful tool to analytically study the multi-level systems with strong transverse couplings.

 \end{abstract}

%\pacs{42.50.Gy, 42.50.Nn, 42.50.Md}
% 42.50.Gy 	Effects of atomic coherence on propagation, absorption, and amplification of light; electromagnetically induced transparency and absorption
%42.50.Nn 	Quantum optical phenomena in absorbing, amplifying, dispersive and conducting media; cooperative phenomena in quantum optical systems
% 42.50.Md 	Optical transient phenomena: quantum beats, photon echo, free-induction decay, dephasings and revivals, optical nutation, and self-induced transparency
\maketitle

%%%%%%%%%%%%%%%%%%%%%%%%%%%%%%%%%%%%%%%%%%%%%%%%%%%%%%%%%%%%%%%%%%%%%%%%%%%%%%%
\section{Introduction}

Strongly driven quantum systems have attracted considerable attention during the past two decades~\cite{PhysRevX.4.031027}. In particular, the development of ultrastrong laser and maser systems opens the doorway for light-matter interactions in the strong- and ultrastrong-coupling regimes. Experimentally, strongly driven two-level systems with a Rabi frequency comparable to or larger than the transition frequency have been realized in a flux qubit~\cite{PhysRevA.95.053824, PhysRevB.89.020503, PhysRevLett.115.133601}, where the driving strength reaches 4.78 GHz, larger than the transition frequency 2.288 GHz. In fact, the strongly driven systems introduce not only many novel phenomena but also the development of fast quantum logic gates, which is of great importance in quantum computation and quantum communication~\cite{PhysRevA.95.062321, PhysRevA.94.012321}. In general, strongly driven systems require strong or ultrastrong coupling between the system and the driving field.

Combined with the modulation, the strongly driven quantum systems exhibit even more interesting phenomena~\cite{Silveri2017Quantum, Grifoni1998Driven, Nori2009Quantum, PhysRevA.94.032323, PhysRevA.96.033802, PhysRevA.89.033812, PhysRevB.92.075312, articlePan, PhysRevLett.123.070505}, such as the coherent destruction of tunneling~\cite{PhysRevLett.67.516}, the driving-induced tunneling oscillations~\cite{PhysRevLett.87.246601}, the suppression of multiphoton resonances~\cite{PhysRevLett.119.053203}, and the localization and transport in a strongly driven two-level system~\cite{PhysRevB.96.014201}. Moreover, periodically toggling a two-level qubit significantly suppresses its decoherence~\cite{PhysRevLett.114.190502, PhysRevB.75.201302, PhysRevB.77.125336} and periodically driving a many-body system offers a powerful tool for coherent manipulation~\cite{PhysRevLett.95.260404, Eckardt2017Atomic, Eckardt2015High}. For these systems, though the time-dependent quantum Hamiltonian can generate a variety of novel phenomena that are inaccessible for ordinary systems, the theoretical challenges arise because many conditional tools such as perturbation theory and rotating-wave approximation can not be directly applied~\cite{Grifoni1998Driven, Ho1985Semiclassical3}, resulting in complex dynamics that is difficult to analyze~\cite{PhysRevB.94.125101}.

Floquet theory is a powerful tool to deal with a periodic time-dependent Hamiltonian, by converting it to an equivalent infinite-dimensional Floquet Hamiltonian in the quasi-level space~\cite{CHU20041, Vleck1929On}. This method has been widely employed in many driven quantum systems, for couplings from weak to strong in either transverse (off-diagonal) or longitudinal (diagonal) form~\cite{PhysRevA.93.033812}. Although numerical results are available by diagonalizing the large-dimensional Floquet Hamiltonian, analytical results under a certain approximation are also desired. The generalized Van Vleck (GVV) nearly degenerate perturbation theory is such an analytical method to reduce the large-dimensional Floquet Hamiltonian to a few dimensional one and to derive the analytical results. In the framework of the Floquet theory, the two-level systems that interact with the external fields via diagonal time-dependent couplings have been extensively explored ~\cite{Ho1984Semiclassical2, Oliver2005Mach,PhysRevLett.87.246601,PhysRevA.75.063414, PhysRevA.79.032301}. For the two-level systems interacting through off-diagonal time-dependent couplings, the combined Floquet and GVV theory was studied only in the weak driving regime~\cite{Chu2004Beyond, Shirley1965Solution, PhysRevA.79.032301}. Despite of its wide applications in the two-level systems, the traditional Floquet and GVV theory has rarely been explored in the three-level systems even in the weak coupling regime~\cite{ Ho1985Semiclassical3}.

In this paper, we present the generalized formalism of double-unitary-transformation (DUT), which make us possible to employ again the GVV perturbation theory to solve the strongly off-diagonal driven systems. Then we give two examples, one for a two-level system, another for a three-level system. Here we focus on the three-level system. We combine Floquet and GVV theory to theoretically investigate a $\Xi$-type three-level system strongly driven by two square-wave modulated and off-diagonal coupling fields, which can induce the experimentally observed modulational diffraction effect in a superconducting qutrit~\cite{HanPRApplied}. Application of the two unitary transformations makes us to use again the GVV perturbation theory and to derive the analytical results. For the numerical calculations, we extend the generalized Floquet formalism to the strongly driven three-level system and obtain nonperturbative results. Comparisons between the analytical and numerical results justify the validity of the predictions from the combined Floquet and GVV theory. This combined theory can be extended to other strongly driven multi-level systems and provides a useful tool to deal with quantum systems in the strongly driven regime.

The paper is organized as follows. In Sec.~\ref{sec:forma}, we present the formalism of DUT. In Sec.~\ref{sec:two}, we present the application of DUT in a two-level system. In Sec.~\ref{sec:three}, combined DUT, Floquet and GVV perturbation theory, we provide the numerical and analytical solutions of a periodically modulated three-level system. We also show the quantum coherence fringes due to the modulation induced diffraction effect in the excited state transition probability. We give conclusions and discussions in Sec.~\ref{sec:con}.

\section{The formalism}
\label{sec:forma}

We propose a double unitary transformation (DUT) scheme to analyze a strongly off-diagonal driven system, so that we can further apply the perturbation theory to the system. In this section, we present the general formalism that illustrate in a simple manner the essence of the DUT. In principle, the time-dependent Hamiltonian of a periodically driven system is
\begin{eqnarray}\label{eq:ghami}
\hat{H}(t)&=&\hat{H}_0+\hat{V}(t),
\end{eqnarray}
where
\begin{eqnarray}\label{eq:period}
\hat{V}(t)=\hat{V}(t+\tau),
\end{eqnarray}
with $\tau$ the modulation period. Here we assume that the time-dependent term $\hat V(t)$ has large off-diagonal elements compared with $\hat{H}_0$, which defines the strong driving regime and makes it difficult to describe the system analytically. Note that the $\hat{V}(t)$ includes not only a large component, but also a small one, i.e.
\begin{equation}
\label{eq:vhami}
\hat{V}(t)=A \hat{V}_L(t)+\xi \hat{V}_S(t),
\end{equation}
with $A\gg \xi$. $\hat{V}_L(t)$ commutes with each other at different times, i.e., $[\hat{V}_L(t),\hat{V}_L(t')]=0$. We use the first unitary transformation to transfer $A \hat{V}_L(t)$ to the diagonal, i.e.,
\begin{eqnarray}
\label{eq:unitary1}
\hat{H}_1(t)&=&\hat{U}_1^\dag H(t)\hat{U}_1\\ \nonumber
&=&\hat{H}'_0+A \hat{V}'_L(t)+\xi \hat{V}'_S(t),
\end{eqnarray}
where $\hat{O}' = \hat{U}_1^\dag \hat O \hat{U}_1$, with $\hat O \in \{\hat H_0, \hat V_L, \hat V_S\}$ and $A \hat{V}'_L(t)$ is diagonal matrix. After the first transformation, the off-diagonal elements of $\hat H_1(t)$ become small. In principle, the unitary matrix $\hat{U}_1$ is system-dependent, which can be realized by having the eigenvectors of the system driven by a strong field. Obviously, the first unitary transformation is not necessary for a strong longitudinal (diagonal) modulation.

The second unitary transform is designed to remove the time-dependence of the diagonal terms in $\hat H_1(t)$~\cite{Coote2017}. This can be realized in the interaction picture by
\begin{equation}\label{eq:unitary2}
\hat{U}_2(t) ={\mathcal T}\exp\left(-i\int_0^tA \hat{V}'_L(t')dt'\right),
\end{equation}
where $\mathcal{T}$ denotes the forward-time ordering and hereafter we set $\hbar=1$. This transformation Eq.~(\ref{eq:unitary2}) has also been employed to estimate the effective tunneling matrix elements of a periodically driven many-body systems~\cite{Eckardt2017Atomic, PhysRevLett.95.260404}. Then the transformed Hamiltonian becomes
\begin{eqnarray}\label{eq:uhami2}
  \hat{H}_2(t)&=&\hat{U}_2^\dag(t)\hat{H}_1(t)\hat{U}_2(t)-i \hat{U}_2^\dag(t)\frac{\partial \hat{U}_2(t)}{\partial t}.\\ \nonumber
\end{eqnarray}
 We require that the Hamiltonian $\hat H_2(t)$ has large and time-independent diagonal terms but small and time-dependent off-diagonal ones. Thus we can define the diagonal terms as the unperturbed part. In fact, similar idea has already been discussed in many-body systems~\cite{Eckardt2015High}. Starting from Eq.~(\ref{eq:uhami2}), we further combine the Floquet theory and the nearly-degenerate perturbation theory, to solve the problem analytically~\cite{Vleck1929On}. Below we illustrate the DUT method in detail by presenting two examples, one for a two-level system and another for a three-level system.

\section{Perturbative solution for a two-level system}
\label{sec:two}

We consider a two-level system with a strong off-diagonal coupling. A strong field couples the two states with a time-dependent Rabi frequency $\varepsilon(t)$. The Hamiltonian is written as
\begin{equation}\label{eq:twoh}
\hat{H}(t)=-\frac{1}{2}\left(
\begin{array}{c c}
-\Delta &\varepsilon(t) \\
 \varepsilon(t) &\Delta
\end{array}
\right),
\end{equation}
where
\begin{equation}
\varepsilon(t)=\varepsilon_0+A\cos(\omega t).
\end{equation}
Here the modulation period is $\tau=2\pi/\omega$ and the detuning is small $\Delta \ll A$. The coupling strength $\varepsilon(t)$ between those two basis states is time-dependent, where the bias consists of a dc component $\varepsilon_0$ and a cosine modulation with a large amplitude $A$ and an angular frequency $\omega$.

By employing a rotation along $y$ direction, the strong off-diagonal elements can be easily transformed to diagonal ones, and thus the first unitary transformation is
\begin{equation}\label{eq:twouni1}
\hat{U}_1=\exp\left(-i \frac{\pi}{4}\hat{\sigma}_y\right),
\end{equation}
where $\hat{\sigma}_y$ is a Pauli matrix. The Hamiltonian after the transformation is
\begin{eqnarray}\label{eq:twohami1}
  \hat{H}_1(t)
  =-\frac{1}{2}\left(
\begin{array}{c c}
\varepsilon(t) &\Delta \\
 \Delta &-\varepsilon(t)
\end{array}
\right),
\end{eqnarray}
which has been studied extensively~\cite{Oliver2005Mach, Shevchenko20101, Shevchenko2012Multiphoton,PhysRevA.79.032301,PhysRevA.81.022117}. Clearly, the first unitary transformation converts a transversely driven system to a longitudinally driven one. Many useful results in the longitudinally driven systems can be applied directly. According to Eq.~(\ref{eq:unitary1}), the time-dependent diagonal matrix with large elements in Eq.~(\ref{eq:twohami1}) is
\begin{eqnarray}\label{eq:largea}
A\hat{V}'_L(t)=\left(
\begin{array}{c c}
-\frac{A}{2}\cos(\omega t) &0 \\
 0 &\frac{A}{2}\cos(\omega t)
\end{array}
\right).
\end{eqnarray}
Then,
\begin{eqnarray}\label{eq:twouni2}
\hat{U}_2(t)&=&{\mathcal T}\exp\left(-i \int_0^t \frac{-A}{2}\cos(\omega t')\hat{\sigma}_z dt'\right)\\ \nonumber
&=&\exp\left(i \frac{A}{2\omega}\sin(\omega t)\hat{\sigma}_z \right).
\end{eqnarray}
The final Hamiltonian after the DUT becomes
\begin{eqnarray}\label{eq:twohami2}
  \hat{H}_2(t)=\frac{1}{2}\left(
\begin{array}{c c}
-\varepsilon_0 &-\Delta \sum\limits_nJ'_{n}e^{-in\omega t} \\\nonumber
\Delta\sum\limits_nJ'_{n}e^{in\omega t}&\varepsilon_0
\end{array}
\right),\\
\end{eqnarray}
where $J'_{n}=J_{n}(A/\omega)$ and we have used
\begin{equation}\label{bessel}
\exp\left(i \frac{A}{\omega}\sin(\omega t)\right)=\sum_{n=-\infty}^{\infty}J'_n e^{i n \omega t}.
\end{equation}
As discussed in Sec.~\ref{sec:forma}, the strong off-diagonal elements with $A$ now turn into a weak off-diagonal elements $J_{\pm} (A/\omega)$ by the DUT, because a larger $A$ give rise to a smaller value of $J_{\pm} (A/\omega)$. This Hamiltonian is exactly the same as in Ref.~\cite{PhysRevA.79.032301}, but the derivation is now much simpler. We can obtain the same analytical solution by further harnessing the Floquet theory and the GVV perturbation theory, as done in Ref.~\cite{PhysRevA.79.032301},
\begin{equation}\label{eq:grat}
\rho_{11}=\sum_{n=-\infty}^{\infty}\frac{1}{2}\frac{(\Delta J'_n)^2}{(\Delta J'_n)^2+(n\omega-\varepsilon_0)^2},
\end{equation}
where $\rho_{11}$ is the time-averaged probability of state $|1\rangle$ with only the first-order perturbation.

\section{Numerical and perturbative solution for a strongly driven three-level system}\label{sec:three}

\begin{figure*}[thp]
%\centering
%\renewcommand{\figurename} {{\bf Figure}}
\includegraphics[width=6.4in]{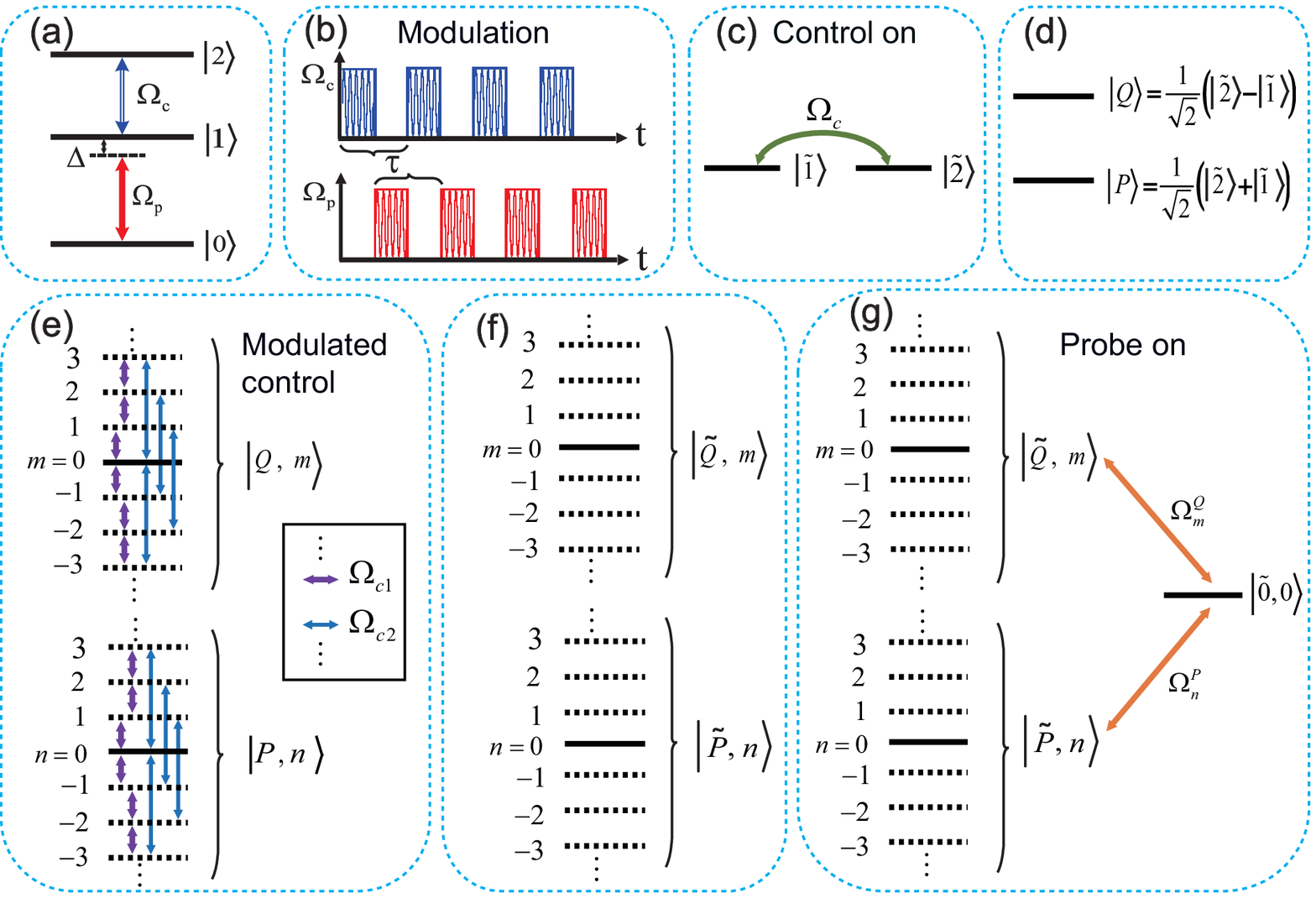}
\caption{\label{fig:pes} (Color online.) Schematic of the quasi-levels of the modulated three-level system. (a) A qutrit with a resonant control-field coupling between states $|1\rangle$ and $|2\rangle$ and with a $\Delta$-detuned probe-field coupling between states $|0\rangle$ and $|1\rangle$. (b) Schematic of modulation, where the control and probe fields are modulated in a complementary square-wave form with a period $\tau$.
A strong control field with Rabi frequency $\Omega_c$ splits, (c) the degenerate dressed states $|\tilde 1\rangle$ and $|\tilde 2\rangle$ into, (d) Autler-Townes doublet $|P\rangle$ and $|Q\rangle$. Under modulation in (b), the state $|P\rangle$ or $|Q\rangle$ further splits into (e), many equally-spaced quasi-levels, which couple in a mirror symmetry within the subspace $|P,n\rangle$ or $|Q,m\rangle$ with $m,n$ being integers. (f) The positions of the quasi-levels are kept untouched after the second unitary transformation, due to the mirror-symmetric coupling. (g) Besides the time-domain interference, the time-domain diffraction appears by sweeping the detuning of the probe field.}
\end{figure*}

As a second example, we consider a generalized $\Xi$-type three-level system shown in Fig.~\ref{fig:pes}(a). A weak probe field couples the ground state $|0\rangle$ and the first excited state $|1\rangle$  with a Rabi frequency $\Omega_p$ and a detuning $\Delta$. A strong control field resonantly couples the excited states $|1\rangle$ and $|2\rangle$ with a Rabi frequency $\Omega_c$. The transition between states $|0\rangle$ and $|2\rangle$ is forbidden. The control and probe fields are complementarily modulated in a square wave form with a period $\tau$ as shown in Fig.~\ref{fig:pes}(b), which are the same as in the double-modulation case studied in Ref.~\cite{HAN2018954}.

\subsection{Numerical solution}
\label{sec:ff}

By decomposing the square-wave modulated fields into many Fourier components, the effective time-dependent Hamiltonian of the modulated three-level system is explicitly given as
\begin{equation}\label{eq:hami}
\hat{H}(t)=\left(
\begin{array}{c c c}
-\frac{\Delta}{2}&-\frac{\Omega_{p}}{4}+\Omega_p(t) &0 \vspace{2mm}\\
 -\frac{\Omega_{p}}{4}+\Omega_p(t) &\frac{\Delta}{2}
&-\frac{\Omega_{c}}{4}+\Omega_c(t)\vspace{2mm}\\
0&-\frac{\Omega_{c}}{4}+\Omega_c(t)&\frac{\Delta}{2}
\end{array}
\right),
\end{equation}
where
\begin{eqnarray}
\Omega_p(t)&=& \sum\limits_{n=1}^{\infty}(-1)^{n}\Omega_{pn}\cos(\omega_{n}t),\\
\Omega_c(t)&=&\sum\limits_{n=1}^{\infty}(-1)^{n+1}\Omega_{cn}\cos(\omega_{n}t),
\end{eqnarray}
with
\begin{eqnarray}
\Omega_{pn}&=&\frac{\Omega_p}{(2n-1)\pi},~~~\Omega_{cn}=\frac{\Omega_c}{(2n-1)\pi},\\
\frac{\omega_{n}}{2\pi}&=&\frac{2n-1}{\tau},~~~n=1,2,3,\cdots
\end{eqnarray}
Clearly, the three-level system is driven by two polychromatic fields and the Rabi frequencies are $\Omega_{pn}$ and $\Omega_{cn}$ for the frequency component $\omega_{n}$. Here we focus on the near- and on-resonance cases. The Hamiltonian in Eq.~(\ref{eq:hami}) is expressed in the dressed-state basis~\cite{API} $|\tilde{\alpha}\rangle,~\alpha=0,1,2$.

The Floquet theory provides an exact formulation for the time-periodic problems, as well as a combined picture for the three-level system and the two fields by using quasi-levels~\cite{Ho1983Semiclassical1}. According to the Fourier expansion theorem~\cite{FS}, the periodic Hamiltonian $\hat{H}(t)$ in Eq.~(\ref{eq:hami}) has Fourier components of $n\omega$ with $\omega=2\pi/\tau$,
\begin{equation}\label{eq:floquet}
\hat{H}(t)=\sum_{n=-\infty}^{\infty}\hat{H}^{[n]}\exp(-i n\omega t),
\end{equation}
where $\hat{H}^{[n]}$ are Fourier coefficients and can be spanned by the Floquet-state nomenclature~\cite{PhysRevA.79.032301, CHU20041}, i.e., $|\tilde{\alpha},n\rangle$ with $\alpha$ being the index of the system and the integer $n\in[-\infty,\infty]$ being the index of the quasi-levels. Then the elements of the infinite-dimensional Floquet matrix $\hat{H}_F$ are defined by~\cite{PhysRevA.79.032301}
\begin{equation}\label{eq:fmatrix}
\langle\alpha n|\hat{H}_F|\beta m\rangle=\hat{H}_{\alpha \beta}^{[n-m]}+n\omega\delta_{\alpha \beta}\delta_{n m},
\end{equation}
and the block structure of $\hat{H}_F$ is
\begin{widetext}
\begin{equation}
\label{eq:lthf}
\renewcommand\arraystretch{1.6}
\hat{H}_F=\left(
\begin{array}{c|c|c|c|c|c|c}
\ddots&\vdots&\vdots&\vdots&\vdots&\vdots&\iddots\\
\hline
 \cdots &\hat{H}^{[0]}-2\omega&\hat{H}^{[-1]}&\hat{H}^{[-2]}&\hat{H}^{[-3]}&\hat{H}^{[-4]}&\cdots\\
  \hline
  \cdots&\hat{H}^{[1]}&\cellcolor{blue!20}\hat{H}^{[0]}-\omega&\cellcolor{blue!20}\hat{H}^{[-1]}&\cellcolor{blue!20}\hat{H}^{[-2]}&\hat{H}^{[-3]}&\cdots\\
  \hline
  \cdots&\hat{H}^{[2]}&\cellcolor{blue!20}\hat{H}^{[1]}&\cellcolor{blue!20}\hat{H}^{[0]}&\cellcolor{blue!20}\hat{H}^{[-1]}&\hat{H}^{[-2]}&\cdots\\
  \hline
  \cdots&\hat{H}^{[3]}&\cellcolor{blue!20}\hat{H}^{[2]}&\cellcolor{blue!20}\hat{H}^{[1]}&\cellcolor{blue!20}\hat{H}^{[0]}+\omega&\hat{H}^{[-1]}&\cdots\\
  \hline
  \cdots&\hat{H}^{[4]}&\hat{H}^{[3]}&\hat{H}^{[2]}&\hat{H}^{[1]}&\hat{H}^{[0]}+2\omega&\cdots\\
  \hline
\iddots&\vdots&\vdots&\vdots&\vdots&\vdots&\ddots
\end{array}
\right).
\end{equation}
\end{widetext}
The Floquet matrix $\hat{H}_F$ is then diagonalized,
\begin{equation}\label{eq:diaf}
\hat{H}_F|\gamma l\rangle=q_{\gamma l}|\gamma l\rangle,
\end{equation}
where $q_{\gamma l}$ is a quasilevel eigenvalue and $|\gamma l\rangle$ the corresponding eigenvector.

\begin{widetext}
According to Eq.~(\ref{eq:lthf}), the Floquet matrix for the time-dependent Hamiltonian in Eq.~(\ref{eq:hami}) is given as follows,
\begin{equation}
\label{eq:thf}
\renewcommand\arraystretch{1.6}
\hat{H}_F=\left(
\begin{array}{c |c c c|c c c|c c c |c}
\ddots&&\vdots&&&\vdots&&&\vdots&&\iddots\\
\hline
   &\frac{-\Delta}{2}-\omega & \frac{-\Omega_{p}}{4}&0&0& \frac{-\Omega_{p1}}{2} &0&0&0&0&\\
  \cdots&\frac{-\Omega_{p}}{4} &\frac{\Delta}{2}-\omega & \frac{-\Omega_{c}}{4}&\frac{-\Omega_{p1}}{2} & 0 &\frac{\Omega_{c1}}{2}&0&0&0&\cdots\\
  &0&\frac{-\Omega_{c}}{4}&\frac{\Delta}{2}-\omega &0&\frac{\Omega_{c1}}{2}&0&0&0&0&\\
  \hline
  &0 & \frac{-\Omega_{p1}}{2} & 0&\frac{-\Delta}{2}& \frac{-\Omega_{p}}{4}&0&0 & \frac{-\Omega_{p1}}{2}&0&\\
  \cdots& \frac{-\Omega_{p1}}{2} & 0 &\frac{\Omega_{c1}}{2}&\frac{-\Omega_{p}}{4} &\frac{\Delta}{2} & \frac{-\Omega_{c}}{4}& \frac{-\Omega_{p1}}{2} & 0 &\frac{\Omega_{c1}}{2}&\cdots\\
  &0&\frac{\Omega_{c1}}{2}&0&0&\frac{-\Omega_{c}}{4}&\frac{\Delta}{2}&0&\frac{\Omega_{c1}}{2}&0&\\
  \hline
  &0&0&0&0 & \frac{-\Omega_{p1}}{2}&0&\frac{-\Delta}{2}+\omega & \frac{-\Omega_{p}}{4}&0&\\
  \cdots&0&0&0& \frac{-\Omega_{p1}}{2} & 0 &\frac{\Omega_{c1}}{2}&\frac{-\Omega_{p}}{4} &\frac{\Delta}{2}+\omega & \frac{-\Omega_{c}}{4}&\cdots\\
  &0&0&0&0&\frac{\Omega_{c1}}{2}&0&0&\frac{-\Omega_{c}}{4}&\frac{\Delta}{2}+\omega&\\
  \hline
\iddots&&\vdots&&&\vdots&&&\vdots&&\ddots
\end{array}
\right)
\renewcommand\arraystretch{1.6}
\begin{array}{c c}
\leftarrow&|\tilde{0},-1\rangle\\
\leftarrow&|\tilde{1},-1\rangle\\
\leftarrow&|\tilde{2},-1\rangle\\
\leftarrow&|\tilde{0},0\rangle\\
\leftarrow&|\tilde{1},0\rangle\\
\leftarrow&|\tilde{2},0\rangle\\
\leftarrow&|\tilde{0},+1\rangle\\
\leftarrow&|\tilde{1},+1\rangle\\
\leftarrow&|\tilde{2},+1\rangle\\,
\end{array}
\end{equation}
\end{widetext}
where we only show the block matrices with light blue background in Eq.~(\ref{eq:lthf}).

For a given initial state $|\alpha, 0\rangle$, by truncating the number of the Floquet blocks with a cutoff number $n_c$ and diagonalizing the matrix numerically, the time-averaged transition probability from $|\alpha\rangle$ to $|\alpha'\rangle$ can be calculated,
\begin{equation}\label{eq:trans}
  \bar{T}_{\alpha\rightarrow\alpha'}=\sum_{n=-n_c}^{n=n_c}\sum_{l=-n_c}^{l=n_c}\sum_{\gamma = \{\tilde{0}, \tilde{1}, \tilde{2}\}}|\langle\alpha', n|{\gamma l}\rangle\langle {\gamma l}|\alpha, 0\rangle|^{2},
\end{equation}
which corresponds to the probability of finding the excited state $|\alpha'\rangle$ of the three-level system in the experiment, i.e., $\rho_{\alpha'\alpha'}$. It is worthwhile to note that no approximation is made in obtaining Eq.~(\ref{eq:thf}), so the numerical results are therefore exact and can be applied to all parameter regimes, with either weak or strong driving.

\subsection{Perturbative solution}
\label{sec:an}

From the Hamiltonian in Eq.~(\ref{eq:thf}), we can see that some off-diagonal terms ($\sim \Omega_c$) may be larger than the diagonal ones ($\sim \Delta, \omega$), in the strong driving regime. The GVV theory apparently fails for large off-diagonal terms. To overcome this difficulty, we transform the strong off-diagonal terms ($\sim \Omega_c$) to the diagonal.

In the case of weak probe field $\Omega_p$, we may neglect at this stage the state $|0\rangle$. The resonantly coupled states $|1\rangle$ and $|2\rangle$ become degenerate in the dressed-state basis (Fig.~\ref{fig:pes}(c)) and further split into the Autler-Townes doublet $|P\rangle$ and $|Q\rangle$ when including the coupling $\Omega_c$ (Fig.~\ref{fig:pes}(d))~\cite{PhysRev.100.703}. Note that here the capital P is different from the lowercase p, which is the index of the probe field. Fortunately, by adopting the coupled dressed-state basis via a proper unitary transformation, i.e., $|\tilde0\rangle$, $|P\rangle$, and $|Q\rangle$ as shown in Fig.~\ref{fig:pes}(d), the first unitary matrix is given by
\begin{equation}\label{eq:threeu1}
\hat{U}_1=\left(
\begin{array}{c c c}
1&0&0 \vspace{2mm}\\
0&\frac{1}{\sqrt{2}}&\frac{-1}{\sqrt{2}}\vspace{2mm}\\
0 &\frac{1}{\sqrt{2}}&\frac{1}{\sqrt{2}}
\end{array}
\right).
\end{equation}
Then the original Hamiltonian in Eq.~(\ref{eq:hami}) becomes
\begin{eqnarray}\label{eq:anyly}
\renewcommand\arraystretch{1.6}
\hat{H}_1(t)
=\left(
\begin{array}{c c c}
-\frac{\Delta}{2} & B(t) & -B(t) \vspace{2mm}\\ \nonumber
B(t)&
\frac{\Delta}{2}-\frac{\Omega_c}{4}+\Omega_c(t)
&0\vspace{2mm}\\
-B(t)&0&\frac{\Delta}{2}+\frac{\Omega_c}{4}-\Omega_c(t)
\end{array}
\right),\\
\end{eqnarray}
where
\begin{equation}
B(t)= -\Omega_p/4\sqrt{2}+\Omega_p(t)/\sqrt{2}.
\end{equation}
According to Eq.~(\ref{eq:unitary1}), the time-dependent diagonal matrix in Eq.~(\ref{eq:anyly}) is
\begin{equation}\label{eq:diag}
  A\hat{V}'_L(t) = \begin{pmatrix}
  0 & 0 & 0 \vspace{2mm}\\
  0 & \Omega_c(t)& 0 \vspace{2mm}\\
  0 & 0 &-\Omega_c(t)\\
\end{pmatrix}.
\end{equation}
Here the large number is $\Omega_c$ (i.e. $A=\Omega_c$).
In Eq.~(\ref{eq:diag}), the strong off-diagonal terms ($\sim \Omega_c$) have been shifted to the diagonal ones and the schematics of Floquet states $|P,n\rangle$ and $|Q,n\rangle$ are shown in Fig.~\ref{fig:pes}(e). We find that there is only internal ``mirror" symmetry couplings between Floquet states $|P,n\rangle$ and $|P,n'\rangle$ (or between $|Q,n\rangle$ and $|Q,n'\rangle$), but without cross couplings between $|P,n\rangle$ and $|Q,n'\rangle$ (see Appendix~\ref{app:threeh1}).

\begin{widetext}
The second unitary transformation can be obtained directly as
\begin{eqnarray}
\renewcommand\arraystretch{1.6}
    \hat{U}_2(t)={\mathcal T}\exp\left(-i\int_0^tA\hat{V}'_L(t')dt'\right)
                =
            \begin{pmatrix}
              1 & 0 & 0 \\
              0 & U_{P}(t) & 0\\
              0 & 0 & U_{Q}(t) \\
            \end{pmatrix},
\end{eqnarray}
where
\begin{eqnarray}
   U_{P}(t) &=&\sum_{n=-\infty}^{\infty}\sum_{l=-\infty}^{\infty}\cdots\sum_{g=-\infty}^{\infty}
  J_{n}\bigg(\frac{-\Omega_{c}}{\omega\pi}\bigg)J_{l}\bigg(\frac{\Omega_{c}}{9\omega\pi}\bigg)\cdots
  J_{g}\bigg(\frac{(-1)^{q}\Omega_{c}}{(2q-1)^{2}\omega\pi}\bigg)e^{i[n+3l+\cdots+(2q-1)g ]\omega t} \nonumber\\
  &=&\sum_{n=-\infty}^{\infty}\sum_{l=-\infty}^{\infty}\cdots\sum_{g=-\infty}^{\infty}
  J_{n-3l-\cdots-(2q-1)g }\bigg(\frac{-\Omega_{c}}{\omega\pi}\bigg)J_{l}\bigg(\frac{\Omega_{c}}{9\omega\pi}\bigg)\cdots
  J_{g}\bigg(\frac{(-1)^{q}\Omega_{c}}{(2q-1)^{2}\omega\pi}\bigg)e^{in\omega t},\nonumber\\
   U_Q(t)&=&\sum_{n=-\infty}^{\infty}\sum_{l=-\infty}^{\infty}\cdots\sum_{g=-\infty}^{\infty}
  J_{n}\bigg(\frac{\Omega_{c}}{\omega\pi}\bigg)J_{l}\bigg(\frac{-\Omega_{c}}{9\omega\pi}\bigg)\cdots
  J_{g}\bigg(\frac{(-1)^{q+1}\Omega_{c}}{(2q-1)^{2}\omega\pi}\bigg)e^{i[n+3l+\cdots+(2q-1)g] \omega t}\nonumber\\
  &=&\sum_{n=-\infty}^{\infty}\sum_{l=-\infty}^{\infty}\cdots\sum_{g=-\infty}^{\infty}
  J_{n-3l-\cdots-(2q-1)g}\bigg(\frac{\Omega_{c}}{\omega\pi}\bigg)J_{l}\bigg(\frac{-\Omega_{c}}{9\omega\pi}\bigg)\cdots
  J_{g}\bigg(\frac{(-1)^{q+1}\Omega_{c}}{(2q-1)^{2}\omega\pi}\bigg)e^{in \omega t}.
\end{eqnarray}
The Hamiltonian $\hat{H}_1(t)$ in the interaction picture becomes
\begin{equation}\label{eq:flom}
\hat{H}_2(t)=
\begin{pmatrix}
-\frac{\Delta}{2} & \sum\limits_{n=-\infty}^{\infty}\Omega _{n}^ {P} e^{i n\omega t} &\sum\limits_{n=-\infty}^{\infty} \Omega _{n}^ {Q} e^{i n\omega t} \vspace{2mm} \\
  \sum\limits_{n=-\infty}^{\infty}\Omega _{n}^ {P} e^{-i n \omega t} & \frac{\Delta}{2}-\frac{\Omega_c}{4} & 0 \vspace{2mm}\\
  \sum\limits_{n=-\infty}^{\infty}\Omega _{n}^ {Q} e^{-i n \omega t} & 0 & \frac{\Delta}{2}+\frac{\Omega_c}{4}\\
\end{pmatrix},
\end{equation}
where
\begin{align}
\label{eq:fmb}
 \Omega_n^{P}=&~\Omega_{p} \sum_{l=-\infty}^{\infty} \cdots \sum_{g=-\infty}^{\infty}\Bigg\{\frac{-1}{4\sqrt{2}}J_{n-3l-\cdots-(2q-1)g}\bigg(\frac{-\Omega_{c}}{\omega\pi}\bigg)
 +\sum_{j=1}^{\infty}\frac{(-1)^j}{2\sqrt{2}(2j-1)\pi}\bigg[J_{n-3l-\cdots-(2q-1)g+(2n+1)}\bigg(\frac{-\Omega_{c}}{\omega\pi}\bigg)   \notag \\
 &+J_{n-3l-\cdots-(2q-1)g-(2n-1)}\bigg(\frac{-\Omega_{c}}{\omega\pi}\bigg)\bigg]\Bigg\}J_{l}\bigg(\frac{\Omega_{c}}{9\omega\pi}\bigg)\cdots J_{g}\bigg(\frac{(-1)^q \Omega_{c}}{(2q-1)^2 \omega\pi}\bigg),    \notag \\
 \Omega_n^{Q}=&~\Omega_{p} \sum_{l=-\infty}^{\infty} \cdots \sum_{g=-\infty}^{\infty}\Bigg\{\frac{1}{4\sqrt{2}}J_{n-3l-\cdots-(2q-1)g}\bigg(\frac{\Omega_{c}}{\omega\pi}\bigg)
 +\sum_{j=1}^{\infty}\frac{(-1)^{j+1}}{2\sqrt{2}(2j-1)\pi}\bigg[J_{n-3l-\cdots-(2q-1)g+(2n+1)}\bigg(\frac{\Omega_{c}}
 {\omega\pi}\bigg) \notag  \\
 &+J_{n-3l-\cdots-(2q-1)g-(2n-1)}\bigg(\frac{\Omega_{c}}{\omega\pi}\bigg)\bigg]\Bigg\}J_{l}\bigg(\frac{-\Omega_{c}}{9\omega\pi}\bigg)\cdots J_{g}\bigg(\frac{(-1)^{q+1} \Omega_{c}}{(2q-1)^2 \omega\pi}\bigg).
 \end{align}

Same as Eq.~(\ref{eq:thf}), the Floquet matrix of Eq.~(\ref{eq:flom}) is
\begin{equation}
\renewcommand\arraystretch{1.5}
\hat{H}'_F=\left(
\begin{array}{c| c c c|c c c|c c c |c}
\label{eq:udfgvv}
\ddots&&\vdots&&&\vdots&&&\vdots&&\iddots\\
\hline
  &\Delta_{-1}^0&\Omega_{0}^{P} &\Omega_{0}^{Q}&0& \Omega_{1}^{P} &\Omega_{1}^{Q}&0&\Omega_{2}^{P} &\Omega_{2}^{Q}\\
  \cdots& \Omega_{0}^{P} &\Delta_{-1}^P&0& \Omega_{-1}^{P} & 0& 0&\Omega_{-2}^{P} &0&0&\cdots\\
  &\Omega_{0}^{Q} &0&\Delta_{-1}^Q&\Omega_{-1}^{Q} &0&0&\Omega_{-2}^{Q} &0&0&\\
  \hline
  &0&\Omega_{-1}^{P} &\Omega_{-1}^{Q} &\Delta_{0}^0&\Omega_{0}^{P} &\Omega_{0}^{Q} &0&\Omega_{1}^{P}&\Omega_{1}^{Q} &\\
  \cdots&\Omega_{1}^{P} &0&0&\Omega_{0}^{P}  &\Delta_{0}^P&0 &\Omega_{-1}^{P} &0&0&\cdots\\
  &\Omega_{1}^{Q} &0&0&\Omega_{0}^{Q} &0&\Delta_{0}^Q&\Omega_{-1}^{Q} &0&0&\\
  \hline
  &0&\Omega_{-2}^{P}&\Omega_{-2}^{Q}&0&\Omega_{-1}^{P}&\Omega_{-1}^{Q}&\Delta_{1}^0&\Omega_{0}^{P}&\Omega_{0}^{Q}&\\
  \cdots&\Omega_{2}^{P}&0&0&\Omega_{1}^{P}&0&0&\Omega_{0}^{P}&\Delta_{1}^{P}&0&\cdots\\
  &\Omega_{2}^{Q}&0&0&\Omega_{1}^{Q}&0&0&\Omega_{0}^{Q}&0&\Delta_{1}^Q&\\
  \hline
  \iddots&&\vdots&&&\vdots&&&\vdots&&\ddots \vspace{2mm}
\end{array}
\right)
\renewcommand\arraystretch{1.5}
\begin{array}{c c}
\leftarrow&|\tilde{0},-1\rangle\\
\leftarrow&|\tilde{P},-1\rangle\\
\leftarrow&|\tilde{Q},-1\rangle\\
\leftarrow&|\tilde{0},0\rangle\\
\leftarrow&|\tilde{P},0\rangle\\
\leftarrow&|\tilde{Q},0\rangle\\
\leftarrow&|\tilde{0},+1\rangle\\
\leftarrow&|\tilde{P},+1\rangle\\
\leftarrow&|\tilde{Q},+1\rangle\\
\end{array}
\end{equation}
where
\begin{equation}
\Delta_{n}^0 =-\frac{\Delta}{2}+n\omega,~~~~~\Delta_{n}^P =\frac{\Delta} {2}-\frac{\Omega_c}{4}+n\omega,~~~~~\Delta_{n}^Q =\frac{\Delta} {2}+\frac{\Omega_c}{4}+n\omega.\nonumber\\
\end{equation}
Here the basis are Floquet states $|\tilde{0},n\rangle$, $|\tilde{P},n\rangle$ and $|\tilde{Q},n\rangle$.
\end{widetext}

Note that compared to Fig.~\ref{fig:pes}(e), in the interaction picture, there is no internal coupling proportional to $\Omega_c$ between the Floquet states $|\tilde{P},n\rangle$ and $|\tilde{P},n'\rangle$ (or between $|\tilde{Q},n\rangle$ and $|\tilde{Q},n'\rangle$) in Eq.~(\ref{eq:udfgvv}), as shown in Fig.~\ref{fig:pes}(f). Also, the energies of the quasi-levels in Fig.~\ref{fig:pes}(f) are the same as in Fig.~\ref{fig:pes}(e). However, the eigenstates are reorganized, which make the couplings between the Floquet states $|\tilde{P},n\rangle$ and $|\tilde{P},n'\rangle$ (or between $|\tilde{Q},n\rangle$ and $|\tilde{Q},n'\rangle$) disappear. From the matrix structure of $\hat{H}'_F$ in Eq.~(\ref{eq:udfgvv}), one sees that $|\tilde{0},0\rangle$ couples weakly to $|\tilde{P},n\rangle$ ($|\tilde{Q},m\rangle$) via an off-diagonal term of $\Omega_n^P$ ($\Omega_m^Q$) as shown in Fig.~\ref{fig:pes}(g). This weak coupling validates the GVV method again and the perturbation parameter is $\Omega_p$. By tuning the frequency of the probe field, the Floquet states $|\tilde{0},0\rangle$ can become nearly degenerate with $|\tilde{P},n\rangle$ (or $|\tilde{Q},m\rangle$), namely, $-\Delta/2\approx\Delta_n^P$ (or $-\Delta/2\approx\Delta_m^Q$). Following the standard GVV method and to the first order, Eq.~(\ref{eq:udfgvv}) is reduced to a $3\times3$ matrix by simply neglecting all other coupling terms,
\begin{equation}\label{eq:rwa}
 \hat{H}_{\rm GRWA}=\left(
 \begin{array}{c c c}
 -\frac{\Delta}{2} &\Omega_{n}^{P}&\Omega_{m}^{Q}\vspace{2mm}\\
 \Omega_{n}^{P}&\Delta_n^P & 0 \vspace{2mm}\\
 \Omega_{m}^{Q}& 0 &\Delta_m^Q
 \end{array}
 \right),
\end{equation}
where the bases are $|\tilde{0},0\rangle$, $|\tilde{P},n\rangle$ and $|\tilde{Q},m\rangle$. Here the results of generalized rotating wave approximation (GRWA) are exactly the same as that of the first-order GVV perturbation. Note that the effective transverse couplings $\Omega_n^P$ and $\Omega_m^Q$ in Eq.~(\ref{eq:fmb}) oscillatingly decrease as $\Omega_c$ increases.

To go beyond the GRWA, we include all coupling channels and keep the second order terms according to the GVV method (see Appendix~\ref{app:gvv}),
\begin{eqnarray}\label{eq:gvv}
 \hat{H}_{\rm GVV}=
 \left(
 \begin{array}{c c c}
 -\frac{\Delta}{2}+\delta_{0} &\Omega_{n}^{P}&\Omega_{m}^{Q}\vspace{2mm}\\
 \Omega_{n}^{P}&\Delta_n^P+\delta_{P} &\delta_{PQ} \vspace{2mm}\\
 \Omega_{m}^{Q}& \delta_{PQ} &\Delta_m^Q+\delta_{Q}
 \end{array}
 \right),
\end{eqnarray}
where
\begin{eqnarray}\label{eq:seccp}
  \delta_{P} &=& \sum_{\begin{subarray}{c} k=-\infty \\ k\neq n\end{subarray}}^{\infty} \frac{(\Omega_{k}^{P})^2}{\Delta-\Omega_c/4+k\omega},\\
  \delta_{Q} &=&\sum_{\begin{subarray}{c} k=-\infty \\ k\neq m\end{subarray}}^{\infty}\frac{(\Omega_{k}^{Q})^2}{\Delta+\Omega_c/4+k\omega},\label{eq:seccq}\\
  \delta_{PQ} &=&\frac{1}{2}\sum_{\begin{subarray}{c} k=-\infty \\ k\neq m\end{subarray}}^{\infty}\frac{\Omega_{k}^{Q}\Omega_{m-n+k}^{P}}{\Delta+\Omega_c/4+k\omega}\nonumber\\ &+&\frac{1}{2}\sum_{\begin{subarray}{c} k=-\infty \\ k\neq n\end{subarray}}^{\infty}\frac{\Omega_{k}^{P}\Omega_{n-m+k}^{Q}}{\Delta-\Omega_c/4+k\omega},\label{eq:seccpq}\\
  \delta_{0} &=& -\delta_{P}-\delta_{Q}\label{eq:secc0},
\end{eqnarray}
which result from the second-order corrections of the non-degenerate quasi-levels. In fact, $\delta_{0,P,Q}$ are the Stark shift and $\delta_{PQ}$ is the effective coupling between the states $|\tilde{P},n\rangle$ and $|\tilde{Q},m\rangle$. All higher-order terms have been neglected.

\begin{figure}[tbh]
%\centering
%\renewcommand{\figurename} {{\bf Figure}}
\includegraphics[width=3.2in]{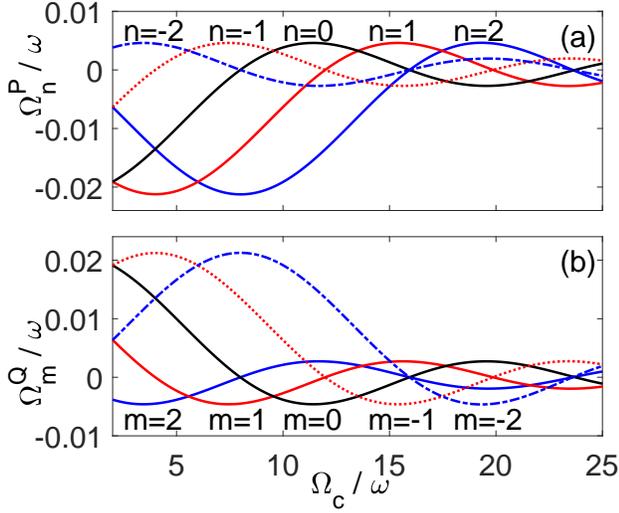}
\caption{\label{fig:omeganpmq} (Color online.) Dependence of $\Omega_n^P$ (a) and $\Omega_m^Q$ (b) on the strong control field coupling strength $\Omega_c$ with a probe field coupling strength $\Omega_p/\omega=0.12$.}
\end{figure}

\begin{figure}[tbh]
%\centering
%\renewcommand{\figurename} {{\bf Figure}}
\includegraphics[width=3.2in]{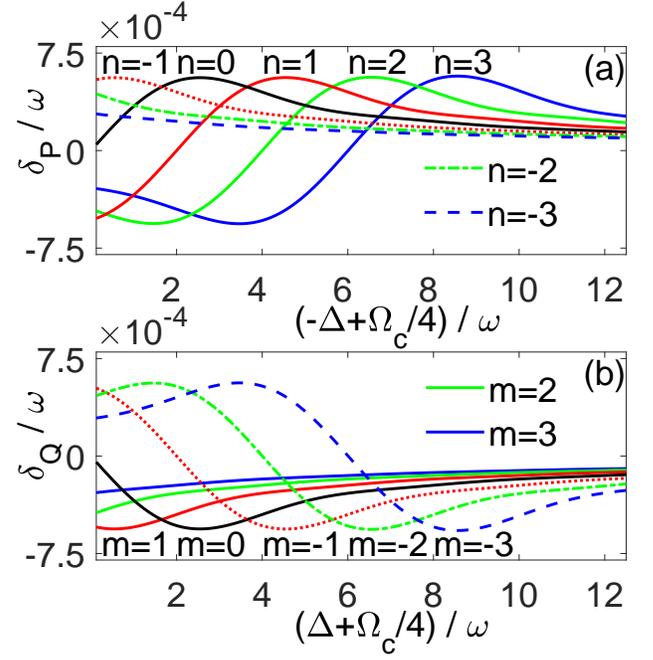}
\caption{\label{fig:deltapandq} (Color online.) (a) Plot of the level shift $\delta_P$ under two-level resonance conditions [i.e., obtained from Eq.~(\ref{eq:seccp}) by setting $\Delta-\Omega_c/4=-n\omega$]. The positions of the two-level resonance in (a) correspond to the bright yellow fringes marked by $n$ in Fig.~\ref{fig:3Drho11}. (b) Same as (a) except for the level shift $\delta_Q$ under two-level resonance conditions [i.e., obtained from Eq.~(\ref{eq:seccq}) by setting $\Delta+\Omega_c/4=-m\omega$]. The positions of the two-level resonance in (b) correspond to the bright yellow fringes marked by $m$ in Fig.~\ref{fig:3Drho11}.}
\end{figure}
\begin{figure}[tbh]
\centering
\includegraphics[width=3.2in]{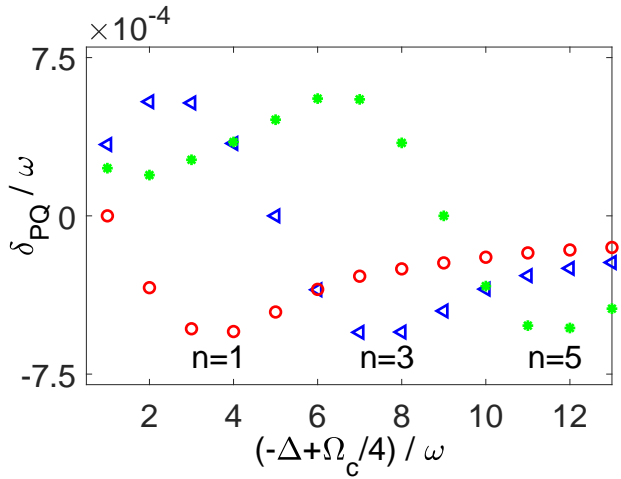}
\caption{\label{fig:deltapq} (Color online.)  Plot of $\delta_{PQ}$ under three-level resonance conditions [i.e., obtained from Eq.~(\ref{eq:seccpq}) by setting $\Delta-\Omega_c/4=-n\omega$ and $\Delta+\Omega_c/4=-m\omega$ simultaneously]. Different from the positions of the two-level resonance, which are the continuous fringes in Fig.~\ref{fig:3Drho11}, the positions of the three-level resonance are the cross points of the fringes in Fig.~\ref{fig:3Drho11}, which are obviously discrete. The probe field coupling strength is $\Omega_p/\omega=0.12$.}
\end{figure}

Let us now investigate the off-diagonal elements of the GVV Hamitonian in Eq.~(\ref{eq:gvv}). Form Eq.~(\ref{eq:fmb}), we notice that $\Omega_n^P$ is proportional to $\Omega_p$ and expressed by a series of Bessel functions with $\Omega_c/\omega$ being the argument. In Fig.~\ref{fig:omeganpmq}, we present $\Omega_n^P$ and $\Omega_m^Q$ as a function of $\Omega_c/\omega$ for different $n$ and $m$ with a fixed parameter $\Omega_p/\omega=0.12$. We observe the translation invariance relation for different n's or m's, i.e.,
\begin{eqnarray}\label{eq:deltanm}
\Omega_n^P\left(\frac{\Omega_c}{\omega}\right)&=&\Omega_0^P\left(\frac{\Omega_c}{\omega}-4n\right),\nonumber\\
\Omega_m^Q\left(\frac{\Omega_c}{\omega}\right)&=&\Omega_0^Q\left(\frac{\Omega_c}{\omega}+4m\right).
\end{eqnarray}
By substituting the resonance conditions $-\Delta/2=\Delta_n^P$ and $-\Delta/2=\Delta_m^Q$, we have $|\Omega_n^P|^2 =|\Omega_m^Q|^2 =|\Omega(\alpha)|^2$, with $\alpha=\Delta\tau/4$, and the function $|\Omega(\alpha)|^2=\sin^2(\alpha)/\alpha^2$ being the diffraction function~\cite{HanPRApplied}.

Figure~\ref{fig:deltapandq} shows the behaviors of the diagonal term $\delta_{P,Q}$ under the same resonance conditions as above. As indicated by Eqs.~(\ref{eq:seccp}) and~(\ref{eq:seccq}), $\delta_{P}$ and $\delta_{Q}$ are proportional to perturbation parameter $\Omega_p^2$, so much smaller than $\Omega_n^P$ and $\Omega_m^Q$ (proportional to $\Omega_p$). Similar to $\Omega_n^P$ and $\Omega_m^Q$, $\delta_{P}$ and $\delta_{Q}$ are also expressed by many Bessel functions and are translationally invariant for different $n$'s and $m$'s. Moreover, Fig.~\ref{fig:deltapq} shows the values of $\delta_{PQ}$ at the three-level resonance points, i.e., $-\Delta/2=\Delta_n^P=\Delta_m^Q$. Similar to $\delta_{P,Q}$, $\delta_{PQ}$ is much smaller than $\Omega_n^P$ and $\Omega_m^Q$.  Note that here we only show the values of $\delta_{P,Q,PQ}$ under resonance conditions, because they become nearly zero at nonresonance points.

The effective GVV Hamiltonian in Eq.~(\ref{eq:gvv}) can be divided into two standard two-level systems (i.e., $|\tilde{0},0\rangle\leftrightarrow|\tilde{P},n\rangle$, and $|\tilde{0},0\rangle\leftrightarrow|\tilde{Q},m\rangle$) except under some special three-level resonance situations. Away from three-level resonances, the time-averaged transition probability from $|0\rangle$ to $|1\rangle$ becomes~\cite{QO, Li2013Motional}
\begin{eqnarray}\label{eq:resgvv}
 \rho_{11}&=&\frac 1 4\sum_{n=-\infty}^{\infty}\frac{(2\Omega_{n}^{P})^2}{(\Delta-\varepsilon_{n}^{P})^2 + (2\Omega_{n}^{P})^2}\nonumber\\
&+&\frac 1 4\sum_{m=-\infty}^{\infty} \frac{(2\Omega_{m}^{Q})^2}{(\Delta-\varepsilon_{m}^{Q})^2 +(2\Omega_{m}^{Q})^2},
\end{eqnarray}
where
\begin{eqnarray}
\varepsilon_{n}^{P}&=&-\Omega_c/4+n\omega+\delta_{P}-\delta_{0},\nonumber\\
\varepsilon_{m}^{Q}&=&\Omega_c/4+m\omega+\delta_{Q}-\delta_{0}.
\end{eqnarray}
Obviously, Eq.~(\ref{eq:resgvv}) contains a series of Lorentzians with each having a peak of 1/4. The peak value of $1/4$ is reasonable because at resonance $\rho_{00}=\rho_{PP}=\rho_{11}+\rho_{22}=1/2$ (or $\rho_{00}=\rho_{QQ}$). Therefore, $\rho_{11}=\rho_{22}=1/4$. Eq.~(\ref{eq:resgvv}) is the main analytical result of this paper. By neglecting the second order terms $\delta_{P, Q, PQ}$, one reaches the GRWA results
\begin{eqnarray}\label{eq:resrwa}
 \rho_{11}&=&\frac 1 4\sum_{n=-\infty}^{\infty}\frac{(2\Omega_{n}^{P})^2}{(\Delta+\Omega_c/4-n\omega)^2 + (2\Omega_{n}^{P})^2}\nonumber\\
&+&\frac 1 4\sum_{m=-\infty}^{\infty} \frac{(2\Omega_{m}^{Q})^2}{(\Delta-\Omega/4-m\omega)^2 +(2\Omega_{m}^{Q})^2}.
\end{eqnarray}
In addition, we have $\rho_{22}=\rho_{11}$, since $|P\rangle=(|1\rangle+|2\rangle)/\sqrt{2}$ and $|Q\rangle=(|2\rangle-|1\rangle)/\sqrt{2}$.

\subsection{Comparisons and discussions}
\label{sec:rd}

\begin{figure}
%\centering
%\renewcommand{\figurename} {{\bf Figure}}
\includegraphics[width=3.2in]{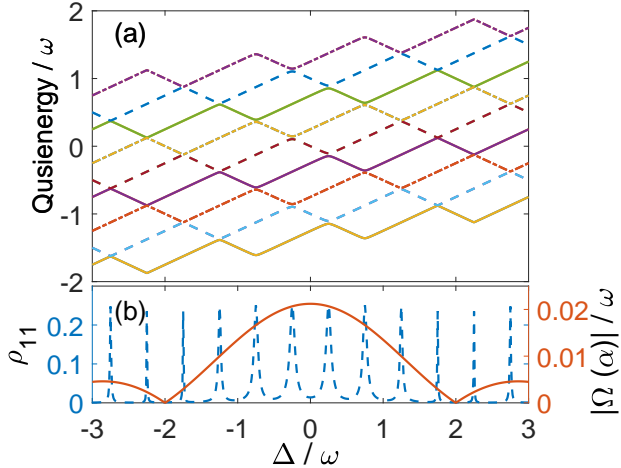}
\caption{\label{fig:quasi} (Color online.) (a) Quasienergies and (b) transition probabilities (blue dashed line, left axis) and diffraction function $|\Omega(\alpha)|$ (orange solid line, right axis) as a function of the detuning $\Delta$ for $\Omega_c/\omega=3$ and $\Omega_p/\omega=0.12$.}
\end{figure}

\begin{figure}[tbh]
%\centering
%\renewcommand{\figurename} {{\bf Figure}}
\includegraphics[width=3.2in]{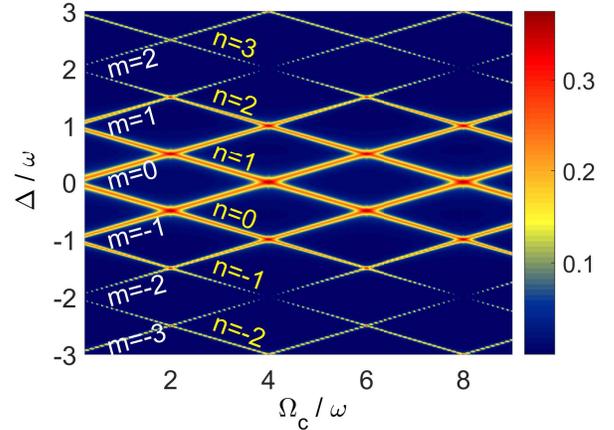}
\caption{\label{fig:3Drho11} (Color online.) Analytical results of $\rho_{11}$ as a function of $\Delta$ and $\Omega_c$ for $\Omega_p/\omega=0.12$ [obtained from Eq.~(\ref{eq:resgvv})].}
\end{figure}

\begin{figure}[tbh]
%\centering
%\renewcommand{\figurename} {{\bf Figure}}
\includegraphics[width=3.2in]{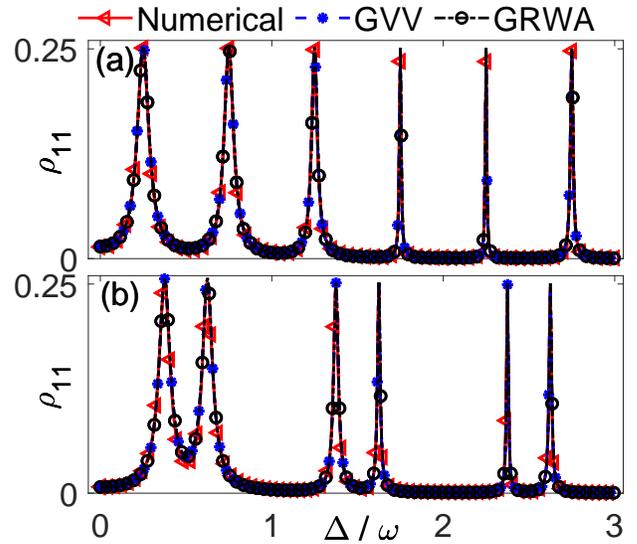}
\caption{\label{fig:grdelta} (Color online.) Comparison of numerical results and analytic GRWA [obtained from Eq.~(\ref{eq:resrwa})] and GVV [obtained from Eq.~(\ref{eq:resgvv})] results of the transition probability $\rho_{11}$ for various $\Delta$ with $\Omega_p/\omega=0.12$, $\Omega_c/\omega=3$ (a) and $\Omega_c/\omega=9.5$ (b).}
\end{figure}

\begin{figure}[tbh]
%\centering
%\renewcommand{\figurename} {{\bf Figure}}
\includegraphics[width=3.2in]{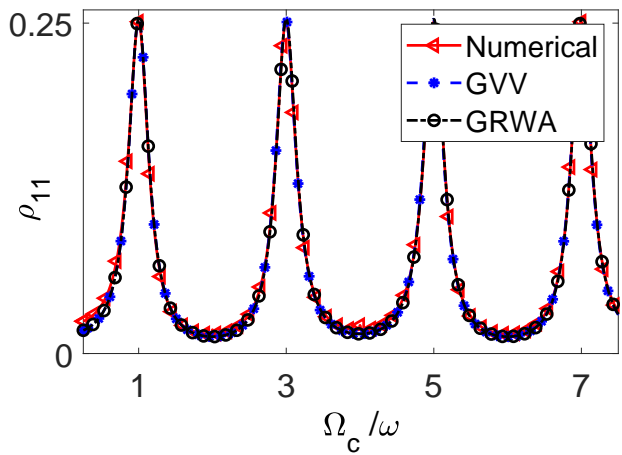}
\caption{\label{fig:gromegac} (Color online.) Same as Fig.~\ref{fig:grdelta} except for various $\Omega_c$ with $\Omega_p/\omega=0.12$ and $\Delta/\omega=0.25$.}
\end{figure}

We plot in Fig.~\ref{fig:quasi} the quasienergies and corresponding time-averaged transition probabilities for $\Omega_c/\omega=3$, computed by truncating the dimension of the Floquet matrix in Eq.~(\ref{eq:thf}) to $n_c=40$ (i.e., Floquet matrix blocks run from -40 to 40). The Floquet states are $|\alpha,n\rangle$ with $\alpha$ the system index and $n\in[-n_c,n_c]$. For $|\alpha,n\rangle$ with a given $n$, the solid lines indicate lower Floquet states, the dashed lines middle Floquet states and the dot-dashed lines upper Floquet states~\cite{PhysRevA.79.032301}. Due to the periodic modulation, the quasienergies exhibit repeated structure by $\omega$. We find some avoided crossings, where the states are strongly mixed and resonant transition between the two anti-crossing levels occurs. These are shown in Fig.~\ref{fig:quasi}(b) by the blue dashed line, which is the time-averaged transition probability of state $|1\rangle$. One sees that the maximal values of the peaks are the same (i.e., 1/4, see also Eq.~(\ref{eq:resgvv})). However, the width of the peaks varies with the increase of $|\Delta|$. In fact, the width is related to the gap of the anti-crossings shown in Fig.~\ref{fig:quasi}(a), and both are determined by the effective coupling $2|\Omega(\alpha)|$, as indicated by Eq.~(\ref{eq:resgvv}). In Fig.~\ref{fig:quasi}(b), we also plot the effective coupling $|\Omega (\alpha)|$ as a function of $\Delta$, and further demonstrate that the width of the peaks broadens as $|\Omega (\alpha)|$ increases. Note that there are zeros in $|\Omega (\alpha)|$, where the resonance ``peaks" disappear.

We show a contour map of transition probability $\rho_{11}$, computed according to Eq.~(\ref{eq:trans}), as a function of $\Delta$ and $\Omega_c$ with $\Omega_p/\omega=0.12$ in Fig.~\ref{fig:3Drho11}. Multiphoton resonance processes occur as the bright yellow fringes shown in the figure. In addition, the peak positions of the fringes $\Delta=\Omega_c/4+m\omega$ and $\Delta=-\Omega_c/4+n\omega$ also indicate that the transitions are multiphoton resonance process~\cite{PhysRevLett.113.247002}. Note that the photon here actually means a quasiphoton with energy $\omega$. We find, at the intersections of the interference fringes, the highest value of $\rho_{11}$ is $1/3$. This is because the three levels of the system are strongly mixed at these intersections. Figure~\ref{fig:3Drho11} for small $\Omega_p$ agrees well with previous experimental results in a superconducting transmon qutrit~\cite{HanPRApplied}. Interestingly, the resonance transitions are suppressed at certain values of $\Delta$ (e.g., $\Delta/\omega=\pm2$), which are similar to the coherent destruction of tunneling in two-level systems~\cite{PhysRevLett.67.516}. In fact, in our modulated three-level system, these destructive interference points correspond to the zeros of the diffraction function as the orange solid line shown in Fig.~\ref{fig:quasi}(b).

In order to justify the validity of our analytic results, we compare the numerical and analytic results by presenting the transition probability of state $|1\rangle$ as a function of $\Delta$ in Fig.~\ref{fig:grdelta}. The numerical solutions are computed by solving the $243\times243$ Floquet matrix in Eq.~(\ref{eq:thf}). The analytical results are obtained by directly solving the $3\times3$ matrix in Eqs.~(\ref{eq:rwa}) and~(\ref{eq:gvv}). Figure~\ref{fig:grdelta}(a) shows the results in the weak control field case of $\Omega_c/\omega=3$ and Fig.~\ref{fig:grdelta}(b) in the strong control field case of $\Omega_c/\omega=9.5$. The higher-order GVV results are not shown since they are almost coincident with the same as the second-order one in the weak probe field regime. The analytic GVV and GRWA results show very good agreement with the numerical solutions in the whole regions we consider. In fact, the difference between the GRWA and GVV is due to the level shift $\delta_{0,P,Q}$. Form Eqs.~(\ref{eq:seccp}) and (\ref{eq:seccq}), one sees that $\delta_{0,P,Q}$ are very small since they all are proportional to the weak probe field $\Omega_p^2$ (see also Fig.~\ref{fig:fig7} in Appendix~\ref{app:expan}). As the strength of the probe field increases, we expect that the GVV shows better fits to exact results than the GRWA (see Fig.~\ref{fig:fig8} in Appendix~\ref{app:expan}).

We further compare the analytical results with the numerical solutions for different control filed $\Omega_c$. In Fig.~\ref{fig:gromegac}, we plot the transition probability $\rho_{11}$ as a function of $\Omega_c$ for a fixed $\Delta/\omega=0.25$. Same as Fig.~\ref{fig:grdelta}, the analytic GRWA and GVV results agree well with the numerical solutions in the whole regime of the figure. Actually, the peak shifts $\delta_{0,P,Q}$ and the peak widthes of the resonance are independent of the control field, since $\Omega(\alpha)$ solely depends on the probe field detuning $\Delta$.

\section{Conclusions}
\label{sec:con}
In summary, we provide a general method to analytically solve strongly coupled two- and three-level systems by a double unitary transformation (DUT) and a combination of the Floquet and GVV perturbation theory. For a periodically modulated three-level system driven by a strong control field, we provide numerical and insightful analytic solutions of the generalized Floquet formalism to explain the quantum interference and diffraction patterns. We extend the generalized Van Vleck perturbation theory to the strong field cases and obtain two analytic solutions, the GRWA and the GVV results. Comparisons show that the two analytic results agree well with the numerical solutions. The general method described here provides a unified theoretical treatment of the modulated two- and three-level systems covering a wide range of parameter space. Applications of the quasilevels to various modulated atomic and artificial atomic systems lead us to a better understanding of the results of spectroscopy measurement and the dynamics of the strongly driven quantum multi-level systems.

\begin{acknowledgements}
We thank J.Q.You and F.Nori for many helpful and intriguing discussions. This work is supported by the National Natural Science Foundation of China under Grants No.~91836101 and No.~11574239. TFL was supported by Science Challenge Project (No. TZ2018003) and BAQIS Research Program (No. Y18G27).
\end{acknowledgements}

\appendix
\begin{widetext}
\section{The Floquet matrix of Eq.~(\ref{eq:diag})}\label{app:threeh1}

The Floquet matrix of Eq.~(\ref{eq:diag}) is given as
\begin{equation}
\renewcommand\arraystretch{1.5}
\hat{V}_F=\left(
\begin{array}{c| c c c|c c c|c c c| c}
\label{eq:udf0}
\ddots&&\vdots&&&\vdots&&&\vdots&&\iddots\\
\hline
  &0&0&0&0&0&0&0&0&0\\
  \cdots&0& 0&0& 0 & \frac{\Omega_{c1}}{2}& 0&0&0&0&\cdots\\
  &0&0&0&0&0&\frac{-\Omega_{c1}}{2}&0&0&0&\\
  \hline
  &0&0&0&0&0&0&0& 0&0&\\
  \cdots&0&\frac{\Omega_{c1}}{2}&0&0 &0 &0 &0 & \frac{\Omega_{c1}}{2}&0&\cdots\\
  &0&0&\frac{-\Omega_{c1}}{2}&0&0&0&0&0&\frac{-\Omega_{c1}}{2}&\\
  \hline
  &0&0&0&0 & 0&0&0&0& 0&\\
  \cdots&0&0&0&0&\frac{\Omega_{c1}}{2} &0&0& 0& 0 &\cdots\\
  &0&0&0&0&0&\frac{-\Omega_{c1}}{2}&0&0&0&\\
\hline
  \iddots&&\vdots&&&\vdots&&&\vdots&&\ddots

\end{array}
\right)
\renewcommand\arraystretch{1.5}
\begin{array}{c c}
\leftarrow&|\tilde{0},-1\rangle\\
\leftarrow&|P,-1\rangle\\
\leftarrow&|Q,-1\rangle\\
\leftarrow&|\tilde{0},0\rangle\\
\leftarrow&|P,0\rangle\\
\leftarrow&|Q,0\rangle\\
\leftarrow&|\tilde{0},+1\rangle\\
\leftarrow&|P,+1\rangle\\
\leftarrow&|Q,+1\rangle\\

\end{array}
\end{equation}

One immediately finds that in Eq.~(\ref{eq:udf0}) the Floquet states $|\tilde{0},n\rangle$ does not couple to any state and thus separates them out. The rest Floquet states $|P,n\rangle$ and $|Q,n\rangle$, as shown in Fig.~\ref{fig:pes}(e), have only internal couplings and the internal coupling is symmetrical, i.e., the coupling between $|P,n\rangle$ and $|P,n-l\rangle$ equals to that between $|P,n\rangle$ and $|P,n+l\rangle$ (the same for $|Q,n\rangle$). Moreover, the coupling is zero for the even $n-n'$ transitions, due to the specific Fourier transform constants of the square wave.
\end{widetext}

\section{Derivation of the $3\times3$ effective Floquet matrix by the GVV theory}
\label{app:gvv}

Our aim is to reduce the infinite-dimensional Floquet matrix in Eq.~(\ref{eq:udfgvv}) into a $3\times3$ effective matrix by the use of GVV perturbation theory~\cite{PhysRevA.32.377, PhysRevA.79.032301,Ho1985Semiclassical3}. Consider the Floquet states $|\tilde{0},0\rangle$ nearly degenerate with $|\tilde{P},n\rangle$ and $|\tilde{Q},m\rangle$. According to the perturbation theory, we expand the $3\times3$ matrix $h$ and its eigenstates $\Phi$ in powers of $\Omega_p$, and the zeroth-order of $\Phi^{(0)}$ is given by
\begin{eqnarray}\label{eq:phi0}
\Phi_0^{(0)}=|\tilde{0},0\rangle,~~
\Phi_P^{(0)}=|\tilde{P},n\rangle,~~
\Phi_Q^{(0)}=|\tilde{Q},m\rangle.
\end{eqnarray}
The zeroth-order $h^{(0)}$ represented by $\Phi^{(0)}$ is
\begin{equation}\label{eq:h0}
 h^{(0)}=\left(
 \begin{array}{c c c}
 -\frac{\Delta}{2}&0&0\vspace{2mm}\\
 0&\Delta_n^P & 0 \vspace{2mm}\\
0& 0 &\Delta_m^Q
 \end{array}
 \right).
\end{equation}
Following the GVV perturbation theory, the higher-order terms are given by
\begin{eqnarray}\label{eq:phi1}
\Phi_0^{(1)}&=&\sum_{\begin{subarray}{c} k=-\infty \\ k\neq n\end{subarray}}^{\infty} \frac{-\Omega_k^P}{\Delta-\Omega_c/4+k\omega}|\tilde{P},k\rangle\\\nonumber
&+&\sum_{\begin{subarray}{c} k=-\infty \\ k\neq m\end{subarray}}^{\infty} \frac{-\Omega_k^Q}{\Delta+\Omega_c/4+k\omega}|\tilde{Q},k\rangle,
\end{eqnarray}
\begin{eqnarray}
\Phi_P^{(1)}=\sum_{\begin{subarray}{c} k=-\infty \\ k\neq n\end{subarray}}^{\infty} \frac{\Omega_k^P}{\Delta-\Omega_c/4+k\omega}|\tilde{0},n-k\rangle,
\end{eqnarray}
\begin{eqnarray}
\Phi_Q^{(1)}=\sum_{\begin{subarray}{c} k=-\infty \\ k\neq m\end{subarray}}^{\infty} \frac{\Omega_k^Q}{\Delta+\Omega_c/4+k\omega}|\tilde{0},m-k\rangle.
\end{eqnarray}
\begin{equation}\label{eq:h1}
h^{(1)}=\langle\Phi^{(0)}|V'|\Phi^{(0)}\rangle
=\left(
 \begin{array}{c c c}
 0&\Omega_{n}^{P}&\Omega_{m}^{Q}\vspace{2mm}\\
 \Omega_{n}^{P}&0& 0 \vspace{2mm}\\
 \Omega_{m}^{Q}& 0 &0
 \end{array}
 \right).
\end{equation}
\begin{eqnarray}\label{eq:h2}
h^{(2)}&=&\langle\Phi^{(0)}|V'|\Phi^{(1)}\rangle-h^{(1)}\langle\Phi^{(0)}|\Phi^{(1)}\rangle\\ \nonumber
&=&\left(
 \begin{array}{c c c}
 \delta_0&0&0\vspace{2mm}\\
 0&\delta_P& \delta_{PQ}\\
 0&\delta_{PQ}&\delta_Q
 \end{array}
 \right),
\end{eqnarray}
Equations~(\ref{eq:h0}) and (\ref{eq:h1}) form the GRWA results in Eqs.~(\ref{eq:rwa}), and Eq.~(\ref{eq:h0}),~(\ref{eq:h1}) and (\ref{eq:h2}) form the GVV results in Eq.~(\ref{eq:gvv}).

\section{A larger probe field}\label{app:expan}
\begin{figure}[tbh]
%\centering
%\renewcommand{\figurename} {{\bf Figure}}
\includegraphics[width=3.2in]{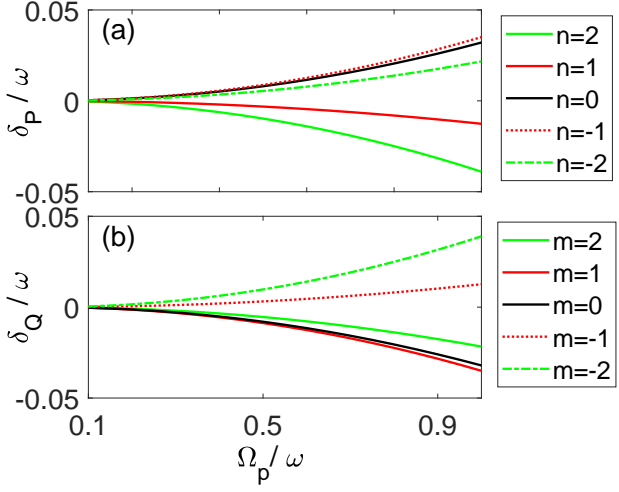}
\caption{\label{fig:fig7} (Color online.) Plots of $\delta_P$ (a) and $\delta_Q$ (b) as a function of $\Omega_p$ under two-level resonance conditions, i.e., obtained from Eq.~(\ref{eq:seccp}) by setting $\Delta-\Omega_c/4=-n\omega$ (a) and from Eq.~(\ref{eq:seccq}) by $\Delta+\Omega_c/4=-m\omega$ (b), respectively, with $\Omega_c/\omega=3$.}
\end{figure}

\begin{figure}[tbh]
%\centering
%\renewcommand{\figurename} {{\bf Figure}}
\includegraphics[width=3.2in]{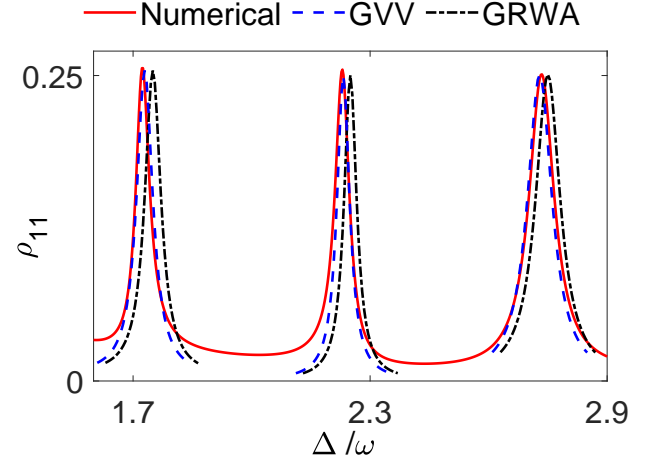}
\caption{\label{fig:fig8} (Color online.) Same as Fig.~\ref{fig:grdelta}(a), except for $\Omega_p/\omega=0.5$.}
\end{figure}

In the periodically driven three-level system, we assume $\Omega_p$ is small as a perturbation parameter. Therefore, the difference between GVV and GRWA (i.e., $|\delta_{P}|$ and $|\delta_{Q}|$) is small. As $\Omega_p$ increases, as shown in Fig.~\ref{fig:fig7}, we observe increasing difference between the two analytical predictions and the effects of $\delta_{P,Q}$ are not negligible. In Fig.~\ref{fig:fig8}, we compare the numerical and analytic results of $\rho_{11}$ from both the GVV and the GRWA for a larger $\Omega_p/\omega=0.5$. One immediately sees that the GVV fits better than the GRWA to the exact numerical results, indicating the deviation of the GRWA and the validity of the GVV.
 \\[1pt]

%\bibliography{ref}

\end{document}